\RequirePackage{lineno}
\documentclass[superscriptaddress,reprint,amsmath,amssymb,prb]{revtex4-1}

\usepackage{graphicx}
\usepackage{dcolumn}
\usepackage{bm}
\usepackage{color,soul}

\usepackage{hyperref}

\usepackage{float}
\usepackage{booktabs} 
\usepackage{subfigure} 
\usepackage[resetlabels]{multibib}

\newcommand{\MI}[1]{\color{black}{#1} \color{black}}

\newcommand{\SK}[1]{\color{black}{#1} \color{black}}

\begin{document}

\title{Elastocaloric signature of nematic fluctuations}
\author{M.\,S.\,Ikeda}
\author{T.\,Worasaran}
\author{E.\,W.\,Rosenberg}
\author{J.\,C.\,Palmstrom}
\affiliation{Geballe Laboratory for Advanced Materials and Department of Applied Physics, Stanford University, Stanford, CA 94305, USA}
\affiliation{Stanford Institute for Materials and Energy Science, SLAC National Accelerator Laboratory, 2575 Sand Hill Road,
Menlo Park, California 94025, USA}
\author{S. A.\ Kivelson}
\affiliation{Stanford Institute for Materials and Energy Science, SLAC National Accelerator Laboratory, 2575 Sand Hill Road,
Menlo Park, California 94025, USA}
\affiliation{Geballe Laboratory for Advanced Materials and Department of Physics, Stanford University, Stanford, CA 94305, USA}
\author{I.\,R.\,Fisher}
\affiliation{Geballe Laboratory for Advanced Materials and Department of Applied Physics, Stanford University, Stanford, CA 94305, USA}
\affiliation{Stanford Institute for Materials and Energy Science, SLAC National Accelerator Laboratory, 2575 Sand Hill Road,
Menlo Park, California 94025, USA}

\date{\today}

\maketitle




\textbf{The elastocaloric effect (ECE) is a thermodynamic quantity relating changes in entropy to changes in strain experienced by a material. As such, ECE measurements can provide valuable information about the entropy landscape proximate to strain-tuned phase transitions. For ordered states that break only point symmetries, bilinear coupling of the order parameter with strain implies that the ECE can also provide a window on fluctuations above the critical temperature, and hence, in principle, can also provide a thermodynamic measure of the associated susceptibility. To demonstrate this, we use the ECE to sensitively reveal the presence of nematic fluctuations in the archetypal Fe-based superconductor Ba(Fe$_{1-x}$Co$_{x}$)$_2$As$_2$. By performing these measurements simultaneously with elastoresistivity in a multimodal fashion, we are able to make a direct and unambiguous comparison of these closely related thermodynamic and transport properties, both of which are sensitive to nematic fluctuations. As a result, we have uncovered an unanticipated doping-dependence of the nemato-elastic coupling and of the magnitude of the scattering of low energy quasi-particles by nematic fluctuations -- while the former weakens, the latter increases dramatically with increasing doping.}



The ECE has been extensively studied in the context of refrigeration, typically utilizing first-order martensitic phase transformations \cite{Rod80.1,Bon08.1,Qia16.1} to induce cooling. Standard ECE techniques that have been so fruitful in this context, however, lack the sensitivity to probe the more subtle entropy effects associated with continuous phase transitions, and consequently the ECE has been largely overlooked in the study of quantum materials and their ordering phenomena. Recently, we have introduced a new AC (i.e. dynamic) ECE technique\cite{Ike19.1,Str20.1} that allows for much more sensitive measurement of the temperature changes induced by oscillating uniaxial stress. This technique not only permits sensitive measurements of the entropy landscape proximate to continuous phase transitions\cite{Ike19.1,Wor20.1,Str20.2} under various thermodynamic conditions including offset strains, magnetic fields etc., but as we will show, for cases where strain couples bilinearly to the order parameter it can also be sensitive to the associated susceptibility far above the phase transition. 

A prominent example of a phase transition for which strain couples bilinearly to the order parameter is that of electronic nematic order. For example, in many tetragonal iron based superconductors, $C_4$ rotational symmetry as well as $\sigma_{\rm x}$ and $\sigma_{\rm y}$ mirror planes are broken at the nematic phase transition, yielding an order parameter with $B_{\rm 2g}$ (i.e. $\rm xy$) symmetry. Application of strain with the same symmetry ($\varepsilon_{\rm B_{2g}} = \varepsilon_{\rm xy}$) via external stresses for temperatures above the critical temperature induces a finite nematic order parameter ($\Psi_{B_{2g}}$) due to this bilinear coupling (see Fig.\,\ref{Fig0}(a)). \MI{Conversely, the coupling also implies that a nematic transition must involve a change in structural symmetry, so that one of the elastic stiffness tensor elements (here $c_{66}$) must vanish upon approach to the critical temperature\cite{Fer10.1,Boe14.1}. Using appropriate Maxwell relations, the elastocaloric effect as a response to $B_{\rm 2g}$ strain is found to be proportional to $\partial c_{66}/\partial T$. The perturbative expression for the effect of the elasto-nematic coupling on $c_{66}$, which is also what emerges from any mean field Landau theory (see Appendix\,\ref{sec:Landau}) is $c_{66} = c_{66}^0 - \lambda^2 \chi$, where $c_{66}^0$ is the bare elastic modulus, $\lambda$ is the coupling between $\Psi_{B_{2g}}$ and $\varepsilon_{\rm B_{2g}}$, and $\chi$ is the associated susceptibility.
\begin{figure*}[ht]
	\centering
		\includegraphics[width=1\textwidth]{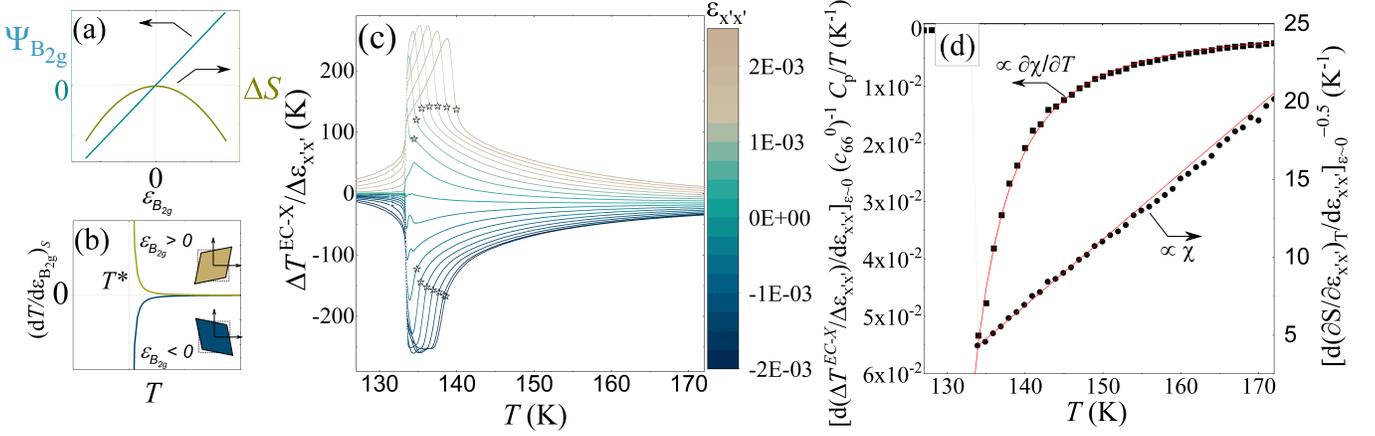}
	\caption{\bf{Entropy and elastocaloric effect associated with nematic fluctuations in iron pnictides}. \rm Panel (a) shows the nematic order parameter $\psi_{B_{2g}}$ as a function of $B_{2g}$ strain, considering only a bi-linear coupling (cyan line, left axis), and the associated entropy (yellow line, right axis), which follows a quadratic strain dependence. The elastocaloric temperature oscillations as a response to $B_{2g}$ strain under adiabatic conditions are sketched in panel (b). They are expected to be proportional to the temperature derivative of the nematic susceptibility $\chi$ and the $B_{2g}$ offset strain. The sign of the effect is thus set by the sign of the applied $\varepsilon_{B_{2g}}$. Panel (c) shows the real part of the experimentally determined elastocaloric temperature oscillations $\Delta{T^{EC-X}}$ of BaFe$_2$As$_2$ as a response to strain along the stress direction  $\varepsilon_{\rm x^\prime x^\prime}$ for various different offset strains. The stress direction $x^\prime$ is aligned along the tetragonal [110] axis (i.e. the primed axis are rotated by 45$^\circ$ in-plane relative to the crystallographic [110] direction). Traces taken under different offset strain conditions are marked using the color map. As discussed in detail in the text, the response observed here is predominantly due to $B_{\rm 2g}$ strain. The magnitude of the elastocaloric temperature oscillations follows for small strains, as expected, a linear relationship with the applied strain $\varepsilon_{\rm x^\prime x^\prime}$ and thus $\varepsilon_{B_{2g}}$ (see Fig.\,\ref{FigS4}). The star symbols show the nematic crossover temperature. Panel(d) shows the linear elastocaloric strain coefficient of BaFe$_2$As$_2$ (square symbols, left axis) as a function of temperature. This quantity is equivalent to $\left[d\left(\partial{S}/\partial{\varepsilon_{\rm x^\prime x^\prime}}\right)_T /d\varepsilon_{\rm x^\prime x^\prime}\right]_{\varepsilon\approx 0}$. It follows our expectations (see Eq.\,\ref{eq4}) for a Curie-Weiss like nematic susceptibility. The nematic transition temperature is marked by the gray dashed line. The square root of the inverse data (circle symbols, right axis) is then proportional to $1/\chi$, which is linear for Curie-Weiss behavior. Our fit, shown as red lines yields a $T^*$ of 124\,K. A similar fit to data corrected for the $B_{2g}$ strain temperature dependence yields a $T^*$ of 117.7\,K (see Appendix\,\ref{sec:Poisson} and Fig.\,\ref{FigS3}).}
	\label{Fig0}
\end{figure*}
Neglecting other sources of strain induced changes in entropy\footnote{Anharmonic effects of the lattice are typically small in solids. Specifically for Co doped BaFe$_2$As$_2$ it has been shown that the lattice part of the Grueneisen parameter is, compared the electronic signatures around the phase transitions, small and weakly temperature dependent\cite{Mei12.1}. The anti-symmetric thermal expansion associated with the effect discussed here is observable in the thermal expansion only through a uniaxially clamped crystal (under thermodynamic conditions comparable to the once applied here) and is expected to be significantly larger than the values observed in the freestanding material.} we find for the quasi-adiabatic\cite{Ike19.1,Str20.1} elastocaloric temperature change as a response to $B_{\rm 2g}$ strain 
\begin{equation}
\begin{split}
    \left(\frac{dT}{d\varepsilon_{B_{2g}}}\right)_S&= -\left(\frac{\partial S}{\partial \varepsilon}\right)_{T,\varepsilon_i}\left(\frac{\partial S}{\partial T}\right)_\varepsilon^{-1}=\\ &=\frac{\partial{c_{66}}}{\partial T}\left(\frac{T}{C_\varepsilon}\right)\varepsilon_{B_{2g}}=\\
    &=-\lambda^2\frac{d\chi}{dT}\left(\frac{T}{C_\varepsilon}\right)\varepsilon_{B_{2g}}.
    \label{eq4}
\end{split}
\end{equation}
Here, $C_\varepsilon$ is the heat capacity at constant strain, which can be adequately approximated \footnote{For solids $C_\varepsilon$, $C_V$, and $C_p$ differ only by a small amount at cryogenic temperatures. This difference is neglected throughout the manuscript} by $C_p$. The subscript in $\varepsilon_i$ denotes that all strain components, except for one, are kept constant.}


  Equation\,\ref{eq4} is remarkable in several ways. First, the sign of the elastocaloric effect due to nematic fluctuations depends on the sign of the offset strain $\varepsilon_{B_{2g}}$. For materials with entropy changes dominated by nematic fluctuations, either a warming or cooling is observed when stress is applied depending on whether the material experiences tensile or compressive offset strain (see Fig.\,\ref{Fig0}(b)). Second, besides the separately measured (and at intermediate temperatures weak) temperature-dependence of $C_V/T$, the temperature-dependence of $\left(dT/d\varepsilon_{B_{2g}}\right)_S$ is determined by $d\chi/dT$, such that integration of the measured ECE signal can yield the nematic susceptibility
  , here from a truly thermodynamic measurement. While the presence of nematic fluctuations in the representative material, Ba(Fe$_{1-x}$Co$_{x}$)$_2$As$_2$, that we consider has been previously deduced from other measurements \cite{Fer10.1,Chu12.1,Gal13.1,Boe14.1}, ECE measurements have several crucial advantages. Not least of these are that they can be performed in a multimodal fashion, allowing direct and unambiguous comparison of this fundamental thermodynamic quantity with other experimental probes such as elastoresistivity, a transport property that is sensitive to nematic fluctuations at the Fermi level.


The measurements presented here were carried out by applying oscillating uniaxial stress to bar-shaped samples along the tetragonal [110] axis. This results in entropy oscillations $(\partial{S}/\partial{\varepsilon_{B_{2g}}})_T$, which, if the stress frequency (here around 30\,Hz) is chosen such that a quasi-adiabatic condition is met, translate into a temperature oscillation. The temperature oscillations ($\Delta{T^{EC}}$) were measured using a thermocouple attached to the center of the sample. Resistivity measurements were performed simultaneously using a standard four point technique and the elastoresistivity extracted using an amplitude demodulation technique\cite{Hri18.1}. The measurements were repeated in the presence of static offset strains introduced by an additional static stress component.


Figure\,\ref{Fig0} (c) illustrates representative data showing the temperature-dependence of the in-phase component of the AC elastocaloric temperature oscillations ($\Delta{T^{EC-X}}$) measured on BaFe$_2$As$_2$. As expected (compare to Fig.\,\ref{Fig0}(b)), a diverging temperature dependence is observed upon cooling towards the nematic critical temperature (134.5\,K), characteristic of a diverging nematic susceptibility.

 The strains that are induced by the [110] stress $\sigma_{\rm x^\prime x^\prime}$ can be decomposed into symmetry-conserving ($\varepsilon_{\rm A_{1g}}$) and symmetry-breaking ($\varepsilon_{\rm B_{2g}}$) strains. As shown through an equivalent experiment applying stress along the tetragonal [100] axis (see Appendix\,\ref{sec:B1g} as well as Figs.\,\ref{FigS1} and \ref{FigS6}), both, $A_{1g}$ and $B_{1g}$ symmetry strains yield significantly smaller elastocaloric signals above the nematic phase transition, such that the response observed here is determined to be predominantly due to $B_{2g}$ symmetry strain.\footnote{We note in passing that within these experiments, $\varepsilon_{\rm x^\prime x^\prime}$ is controlled instead of $\varepsilon_{B_{2g}}$. This is a different thermodynamic condition yielding a slightly more complicated, but ultimately equivalent, expression for the elastocaloric temperature oscillations as a response to oscillating strain $\varepsilon_{\rm x^\prime x^\prime}$ (see Appendix\,\ref{sec:Landau}).} 

The linear strain coefficient of the ECE can be extracted by taking iso-thermal cuts through the temperature sweep data shown in Fig.\,\ref{Fig0}(c). Multiplied by the separately measured heat capacity (see Appendices\,\ref{sec:strainsweep} and \ref{sec:HC}), and divided by $T$, and suitably corrected for the $T$-dependence of the Poisson ratio (see Appendix\,\ref{sec:Poisson}), this coefficient then directly reveals the $T$-dependence of $\partial\chi/\partial T$ (see Fig.\,\ref{Fig0}(d)). As expected, $\partial\chi/\partial T$ is well described\footnote{For the fit, the expression is multiplied by a factor of 0.77 to account for the fact \cite{Ike19.1} that only a fraction of the total elastocaloric temperature oscillations are measured. In addition, to get the best estimation for $T^*$, the data is also corrected for the temperature dependence of $B_{2g}$ strain.} by a power law. 
Since the nemato-elastic coupling $\lambda$ determines the renormalized nematic transition temperature ($T_S=T^*+\lambda^2/ \left(a_0  c_{66}^0\right)$), the bare electronic transition temperature $T^*$ appears in the numerator as well as in the denominator of Eq.\,\ref{eq4} 
$\left( \lambda^2\frac{\partial\chi}{\partial T} =-c_{66}^0\frac{(T_S-T^*)}{\left(T-T^*\right)^2}\right)$, as long as $\chi$ follows Curie-Weiss behavior (i.e. $\chi=\left(a_0 (T-T^*)\right)^{-1}$. Hence, ECE measurements provide both a very sensitive test of the $T$-dependence of $\chi$ and an excellent means to estimate\footnote{As pointed out earlier\cite{Boe14.1}, the bare elastic modulus $c_{66}^0$ is weakly temperature dependent. This effect is neglected here.} the bare nematic critical temperature $T^*$. A power law fit (red line in Fig.\,\ref{Fig0}(d) and Appendix\,\ref{sec:Poisson}) according to $(T_S-T^{*})/(T-T^*)^{2}$, results in a $T^*$ of 117.7\,K, roughly 17\,K below the nematic phase transition, comparable to what has previously been deduced from other measurements\cite{Chu12.1,Gal13.1,Boe14.1}.
While $\chi(T)$ for BaFe$_2$As$_2$ is very well described by a power law, as we will show shortly, this is not the case for doped samples close to optimal doping.

\begin{figure}
\centering
   \includegraphics[width=0.45\textwidth]{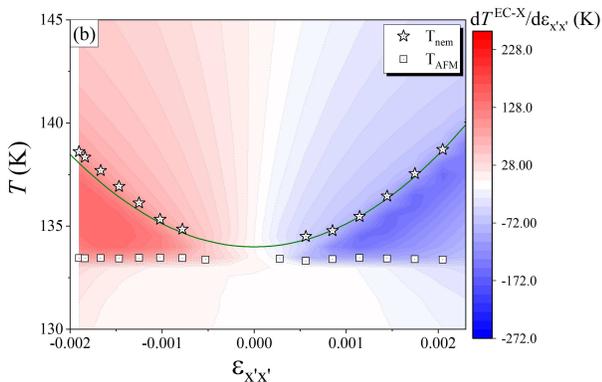}
\caption{\bf{Elastocaloric effect of nematic fluctuations in BaFe$_2$As$_2$}. \rm  In-phase elastocaloric effect data of BaFe$_2$As$_2$ in the temperature-strain plane shown as a color map. The sign change of the ECE as a response to switching from compressive to tensile offset strain, as well as the features identified in the data are clearly visible. The stars correspond to the nematic cross-over temperature, the green line shows a quadratic fit to the data, the squares are attributed to the antiferromagnetic phase transition. The sign change of the ECE as a response to a changing sign of the applied offset strain together with a diverging temperature dependence are the hallmark of nematic fluctuations.}
   \label{Fig1}
\end{figure}

 Figure\,\ref{Fig1} summarizes the elastocaloric effect data on BaFe$_2$As$_2$ in the strain-temperature plane. Closely examining the in-phase part\footnote{Out-of-phase ECE data ($\Delta{T^{EC-Y}}$) is shown and discussed in Appendix\,\ref{sec:OutofPhase}.} and its temperature derivative, an inflection point can be identified (marked by the star symbols in Figs.\,\ref{Fig0}(c) and \ref{Fig1}) below which the data can no longer be described by a $\left(T-T^*\right)^{-2}$ behavior. We interpret the temperature of this inflection point as the cross over temperature for the nematic ``transition'' in the presence of the symmetry breaking offset strain.  Remarkably, a quadratic fit\footnote{Note that at the nematic phase transition $c_{66}$ is expected to be zero, such that the strain transfer ratio should be one and $\varepsilon_{\rm x^\prime x^\prime}=\varepsilon_{B_{2g}}$.} (green line in Fig.\,\ref{Fig1}) reveals an extremely large value for the normalized biquadratic coupling constant $\gamma/a_0$ of 121.1\,K/\%$^2$. Thus, for larger strains and temperatures close to $T^*$, the biquadratic term in the Landau free energy ($\frac{\gamma}{2} \Psi_{B_{2g}}^2\varepsilon_{B_{2g}}^2$) can not be neglected in the elastocaloric effect (see Appendix\,\ref{sec:strainsweep}). Below another characteristic temperature, the in-phase elastocaloric effect reduces significantly in magnitude. This feature 
 coincides with the antiferromagnetic phase transition (squares in Fig.\,\ref{Fig1} and dashed line in Fig.\,\ref{Fig0} (c)), %
 and is not strongly dependent on the magnitude of the offset strain reflecting the fact that below $T_S$ the small offset strain merely changes the domain population in the sample and hence the strain experienced within each nematic domain is independent of changes in the length of the sample.
 
The ECE measurements described above are especially revealing  when compared to elastoresistivity measurements that are performed simultaneously. This comparison allows us to draw conclusions about the nemato-elastic coupling 
and the nematic scattering cross section. While according to Eq.\,\ref{eq4} the elastocaloric effect is proportional to $d\chi/dT$, the antisymmetric part of the strain-induced change in resistivity $\left(\frac{\Delta\rho}{\rho}\right)_{B_{2g}}$ is given by:
\begin{equation}
   m_{B_{2g}}^{B_{2g}}=\frac{d\left(\frac{\Delta\rho}{\rho}\right)_{B_{2g}}}{d\varepsilon_{B_{2g}}}= g_{x,T} \lambda_x \chi,
    \label{eq7}
\end{equation}
where $g_{x,T}$ is a proportionality constant and $\lambda_x$ is the previously defined nemato-elastic coupling, which we now explicitly allow to vary with composition, $x$. Thus if $g_{x,T}$ is temperature-independent (i.e. $g_{x,T}=g_{x}$), then the normalized elastocaloric temperature oscillations as a response to $B_{2g}$ strain are expected to be linearly proportional\footnote{Under small offset strains the measured longitudinal elastoresistivity is dominated by $m_{B_{2g}}^{B_{2g}}$} to the temperature derivative of $m_{B_{2g}}^{B_{2g}}$. Put another way, comparison of ECE and ER data (described by equations \ref{eq4} and \ref{eq7} respectively) provide a direct window on the doping and temperature dependence of $g_x/\lambda_x$.

\begin{figure}[hb!]
\centering
		\includegraphics[height=0.45\textwidth]{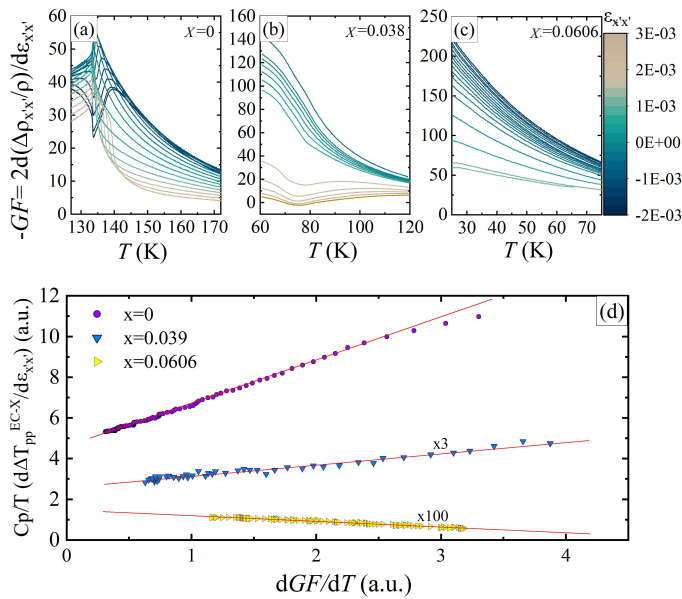}

\caption{
   \label{Fig2}
\bf{Comparison of elastoresistivity and elastocaloric effect for BaFe$_2$As$_2$.}\rm  Panel (a), (b), and (c) show the temperature-dependence of the longitudinal elastoresistivity of Ba(Fe$_{1-x}$Co$_{x}$)$_2$As$_2$ with $x=$0, 0.0387, and 0.0606, respectively. The elastoresistivity arises from oscillating uniaxial stress in the x-direction (which is aligned with the tetragonal [110] axis), while controlling for constant $\varepsilon_{\rm x^\prime x^\prime}$ (values indicated in legend). Traces taken under different offset strain conditions are marked using the color map. Without non-linear effects in the elastoresistivity \cite{Pal17.1} (discussed in the main text), the data  would collapse onto a single trace in each panel. Note that the resistivity as a function of static offset strain is not symmetric with respect to the strain neutral point\cite{San20.1} such that non linearities affect the data differently at compressive and tensile strains. Panel (d) shows the magnitude of the elastocaloric temperature oscillations multiplied by heat capacity over temperature (which is then equivalent to $-\left(\partial{S}/\partial{\varepsilon_{\rm x^\prime x^\prime}}\right)_T$) as a function of the temperature derivative of the longitudinal elastoresistivity for $x=$0, 0.0387, and 0.0606. The data for $x=$0.0387 and 0.0606 is multiplied by factor of 3 and 100, respectively. Although different offset strains were used for each doping (see Appendix\,\ref{sec:Landau} and Figure\,\ref{FigS5}), here, for each composition, data closest to the strain neutral point is shown, for which non-linear effects are expected to be smallest. Temperature is an implicit variable in panel (d) and spans a range of about 40\,K above the nematic transition temperature. The elastoresistance and elastocaloric effect measurements were performed simultaneously, for identical strain conditions. As discussed in the main text and Appendix\,\ref{sec:Landau}, a linear relation is expected for small strains as long as $g_{x,T}$, connecting the elastoresistivity and the nematic susceptibility, does not show a strong temperature dependence itself. The linear fits (red lines in panel (d)) to traces taken under small offset strains confirm that this is indeed the case.}
\end{figure}

Figure\,\ref{Fig2}(a), (b), and (c) show the measured longitudinal elastoresistivity
$2d\left(\frac{\Delta \rho}{\rho_{0}}\right)_{\rm x^\prime x^\prime}/d\varepsilon_{\rm x^\prime x^\prime}$ (i.e. elastoresistive gauge factor $GF$)
of Ba(Fe$_{1-x}$Co$_{x}$)$_2$As$_2$ with $x$=0, 0.0387, and 0.0606 for uniaxial stress applied along the tetragonal [110] axis (x'-direction), determined simultaneously to the elastocaloric effect. Even for the $x=0$ parent compound 
shown in panel (a), large non-linear effects \cite{Pal17.1} are observed in the elastoresistivity, without which all traces would collapse onto a single curve. Also, the temperature dependence of the longitudinal elastoresistivity deviates significantly from published $m_{B_{2g}}^{B_{2g}}$ elastoresistivity data \cite{Kuo16.1,Pal20.1}, since due to the softening of $c_{66}$ the in-plane Poisson ratio $\nu$ is temperature dependent, and thus the ratio of symmetry conserving $A_{1g}$ and symmetry breaking $B_{2g}$ strains changes during temperature sweeps. Identifying the trace at smallest offset strain (the zero strain point) from the elastocaloric effect data (see Appendix\,\ref{sec:zeroStrain} and Fig.\,\ref{FigS7}), we extract the temperature dependence of the in-plane Poisson ratio by comparing published elastoresistivity data with the longitudinal elastoresistivity from the present measurements. This estimation agrees well (see Fig.\,\ref{FigS2} )  with a calculated Poisson ratio considering a full softening of $c_{66}$.
Significantly, and irrespective of this correction, the strain condition during the elastoresistivity and the elastocaloric effect measurements is the same for each offset strain trace, respectively. It is thus possible to directly compare both measurements.  This is what is shown in Fig.\,\ref{Fig2}(d), in which we plot the elastocaloric temperature oscillations multiplied by $C_p/T$ versus the temperature derivative of the measured longitudinal elastoresistivity. For each composition, the data taken at smallest offset strain is shown, for which a linear dependence is observed from high temperatures all the way down to the nematic transition temperature. For larger offset strains (see Appendix\,\ref{sec:Landau}), linearity is, as expected, only observed in the regime of relatively small susceptibility (i.e. away from the nematic crossover temperature), where non-linear effects in the elastoresistivity are comparably weak. It is thus evident that $g_{x,T}$ does not itself show a strong temperature dependence (i.e $g_{x,T}\approx g_{x}$). Hence, the sub-Curie-Weiss temperature dependence observed earlier in various iron-pnictide superconductors \cite{Kuo16.1} is due to a deviation of the nematic susceptibility itself from Curie-Weiss behavior. As discussed earlier by H.-H.\,Kuo et al.\cite{Kuo16.1}, deviations of the nematic susceptibility from Curie-Weiss behavior due to weak disorder are expected upon approaching both finite and zero temperature phase transitions. The effects are, however, expected to be stronger for quantum critical scenarios. This is particularly interesting considering the doping dependence of the elastocaloric effect discussed next.

\begin{figure*}[ht]
\centering
		\includegraphics[width=0.9\textwidth]{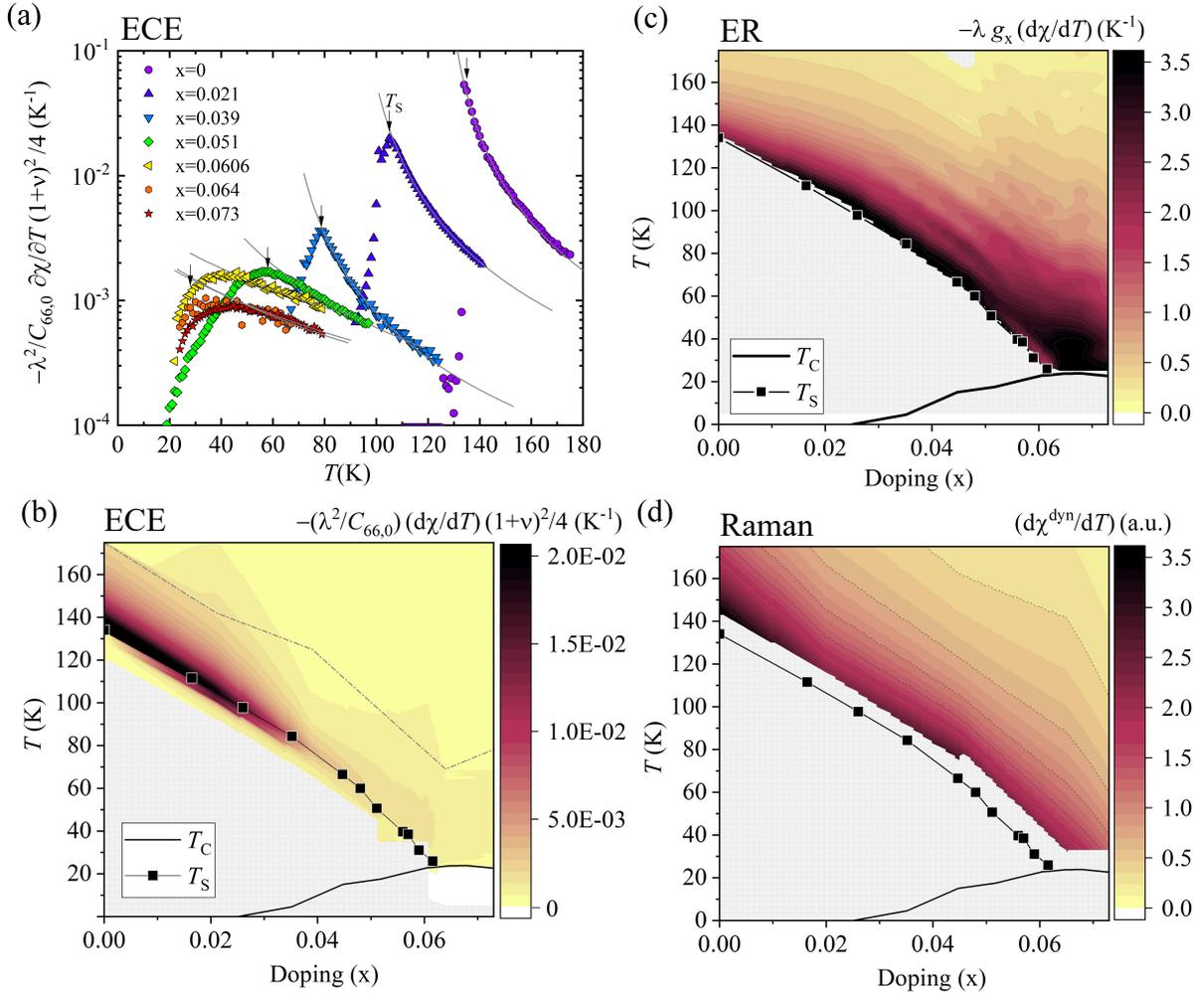}

\caption{
   \label{Fig3}
\bf{Comparison of the temperature and doping-dependence of the elastocaloric effect, elastoresistivity, and Raman data for  Ba(Fe$_{1-x}$Co$_{x}$)$_2$As$_2$.}\rm  Panel (a) shows the linear strain coefficient of the elastocaloric temperature oscillations observed for various Co dopings $x$. For strongly underdoped compositions, the nematic susceptibility is found to follow Curie-Weiss behaviour down to the nematic transition temperature marked using the black arrows. The gray lines correspond to power law fits assuming Curie-Weiss behavior for $\chi$ as discussed in the main text. For higher dopings a deviation from Curie-Weiss behavior is obvious. The magnitude of the effect is compared best in the colorplot shown in panel (b). The data shown in panel (a) is interpolated in parallel to the nematic phase transition line and extrapolated to high temperatures using the fits shown as gray lines in panel (a). The region above the dash-dotted gray line in panel (b) shows extrapolated data. Panel (c) shows the temperature derivative of $m_{B_{2g}}^{B_{2g}}$ elastoresistivity data\cite{Kuo16.1,Pal20.1}. Panel (d) shows the temperature derivative of published \cite{Gal13.1,Gal16.1} Raman dynamical susceptibility data. A weak composition dependence of $\partial{\chi^{dyn}}/\partial{T}$ revealed by Raman scattering has also been observed by Kretzschmar et al.\cite{Kre16.1}.
}
\end{figure*}
We finally compare and analyse our elastocaloric effect results on Ba(Fe$_{1-x}$Co$_{x}$)$_2$As$_2$ with $x=0, 0.021, 0.0387, 0.0512, 0.0606, 0.064,$ and $0.073$.
Figure\,\ref{Fig3} (a) summarizes the linear strain coefficient of the elatocaloric effect for all Co dopings $x$ measured here. The structural transition temperature is marked by black arrows. Inspection of these data reveal that they follow the derivative of a Curie-Weiss form only for Co concentrations smaller than 4\%. For higher dopings a clear departure from Curie-Weiss behavior is observed. Interestingly, this regime agrees well with the doping range for which power law behavior of the nematic transition temperature is observed  -- a characteristic feature of quantum criticality (see Ref.\,\onlinecite{Wor20.1}) -- indicating that the stronger deviations from Curie-Weiss behavior might indeed occur in a quantum critical regime.


In addition to the onset of sub-Curie-Weiss behavior, increasing the Co conentration also results in a significant reduction in the \textit{magnitude} of the elastocaloric effect. It is instructive to plot the same data in a color plot (Fig.\,\ref{Fig3}(b)) and compare it to elastoresistivity (Fig.\,\ref{Fig3}(c)) and an estimate of the dynamic nematic susceptibility $\chi^{dyn}$ obtained from Raman scattering data (Fig.\,\ref{Fig3}(d)). Analysing the magnitude of our elastocaloric effect data, it is important to note that for coordinates along a line close to and parallel to the nematic phase transition line, the magnitude of the observed AC elastocaloric temperature oscillations is close to its value expected from Eq.\,\ref{eq4} since $\varepsilon_{\rm x^\prime x^\prime}\approx\varepsilon_{B_{2g}}$ 
for $T$ near $T_S$. Walking along the nematic phase transition line (for which $\nu\approx1$), the magnitude of the strain response of the AC elastocaloric effect shrinks by a factor of about 35 as the Co concentration increases from $x=$ 0\% to 6.06\%. Considering only the ECE data, this effect could be caused by an $x$ dependent change in $d\chi/dT$ and/or by a weakening of the nemato-elastic coupling $\lambda_x$. To estimate the doping dependence of $d\chi/dT$ we revisit dynamic nematic susceptibility data $\chi^{dyn}$ determined via Raman scattering measurements and published earlier by Y.\,Gallais et al.\cite{Gal13.1}. Figure\,\ref{Fig3}(d) shows the temperature derivative of this data in a color plot. Assuming the nematic susceptibility is well described by the dynamical susceptibility determined from Raman scattering experiments, we find that $d\chi/dT$ drops by at most a factor of 1.5 
over this range of $x$. 
This in turn implies that 
$\lambda_x$ 
must decrease by a factor of about 5 to explain the strong drop in the $B_{2g}$ elastocaloric strain response, which is proportional to $\lambda_x^2$ (see Fig.\,\ref{Fig3}(a)).
The simultaneous dramatic increase in the elastoresistivity upon approaching the putative quantum critical point is then revealed to be due to the doping dependence of $g_x$, which 
increases by approximately one order of magnitude over the same range of $x$. 
In other words, the coupling between quasiparticles at the Fermi energy and the nematic order parameter increases dramatically as the critical temperature of the nematic phase transition is suppressed towards zero temperature. This conclusion can be drawn even in the absence of Raman data since the comparison of elastoresistivity and elastocaloric effcet reveals 
a strongly increasing ratio $g_x/\lambda_x$ as the Co concentration increases towards optimally doped BaFe$_2$As$_2$.  This last point is also in line with results of a recent investigation\cite{San20.1} of the spontaneous elastoresistivity in the ordered phase.

\SK{It remains unclear {\it why} the nemato-elastic coupling decreases so dramatically with doping, and whether this is a ubiquitous effect - possibly associated with universal physics in the proximity of the nematic QCP - or rather some special feature of a specific family of materials.  Independent of the origin of the effect, wherever it occurs, it has interesting implications for the mechanism of superconductivity.  Both on the basis of perturbative analysis\cite{Led15.1} and the results of quantum Monte-Carlo studies\cite{Led17.1}, it is well established that nematic fluctuations greatly enhance superconducting pairing upon approach to a nematic QCP.  However, it has also been shown that elasto-nematic coupling suppresses the relevant critical fluctuations\cite{Lab17.1}, and can essentially eliminate this effect.  In this regard, the empirical observation that $\lambda_x$ is small near the QCP opens a window for a related enhancement of $T_c$.  The fact that $g_x$ is anomalously large near the QCP (also an empirical observation whose explanation is currently unclear) further encourages this idea, as it implies that the low-energy quasi-particle degrees of freedom are strongly coupled to the fluctuating nematic order. Moreover, on the basis of the previously noted\cite{Kuo16.1} fact that in several families of Fe-based superconductors, optimal $T_c$ occurs at a doping value close to that at which $T^\star$ vanishes - i.e. near a putative nematic QCP - it is tempting to speculate that the same physics may apply more broadly.}

\noindent{\bf Materials and methods}\\
\noindent{\bf Sample preparation} \hspace{0.1cm} \\
Bulk single crystal samples were grown by using a self flux technique described in detail in J.-H. Chu et al. \cite{Chu09.1}. The Co-doping was measured for all material batches using electron microprobe analysis (EMPA). The parent compound BaFe$_2$As$_2$ and cobalt metal were used for calibration. Doping variation within a sample and within a batch were found to be characterized by a standard deviation of less than 0.2\%. Bulk single crystals were cleaved into bar shaped samples with typical dimension of 2000x500x50$\rm\mu$m such that the tetragonal [110] axis was aligned with the long side of the bar. \\

\noindent{\bf Uniaxial stress elastocaloric effect measurements} \hspace{0.1cm} \\
Strain is applied to the sample using the commercially available uniaxial stress apparatus CS100 from \textit{Razorbill instruments}. The CS100 cell is designed to compensate for the thermal expansion of the lead zirconate titanate (PZT) stacks \cite{Hic14.1}. Furthermore, due to matching of the thermal expansion of Ba(Fe$_{1-x}$Co$_x$)$_2$As$_2$ (Ref.\,\onlinecite{Dal09.1}) and the cell body/ the sample mounting plates (titanium) \cite{Ike18.1}, the externally applied strain on the sample is almost perfectly independent of temperature for a fixed voltage applied to the PZT stacks. Uniaxial stress is applied along bar shaped samples with a typical dimension of 2000x500x50$\rm\mu$m by affixing it in between two mounting plates that are pushed together/pulled apart using voltage controlled PZT stacks. The sample typically spans a gap of 850$\rm\mu$m. For the uniaxial stress experiment the crystallographic [110] direction is aligned along the direction of the stress $\sigma_{\rm x^\prime x^\prime}$. While the inner PZT stack is used to apply constant DC offset strain, the outer stacks are used in parallel to apply oscillating strain optionally on top of additional DC offset strain. The outer stacks are controlled by the oscillator output voltage of a \textit{Stanford Research} SR860 Lock-In amplifier (whose input is used for detecting the elastocaloric temperature oscillation) amplified by a factor of 25 by a circuit using a \textit{Tegam} 2350 precision power amplifier. The inner stack is controlled by a DC voltage supplied by an auxiliary output of the SR860 amplified by a factor of 25 using a \textit{piezosystem jena} SVR 350-1 bipolar voltage amplifier. The CS100 cell is equipped with a capacitive displacement sensor, measuring the distance change between the mounting plates. The externally applied nominal strain in the sample along the stress axis ($\varepsilon_{\rm x^\prime x^\prime}$) is then determined by scaling the measured length change by the zero volt length of the sample between the mounting plates. The AC strain magnitude is inferred from strain per volt data and the applied AC voltage amplitude. The oscillating strain portion induces temperature oscillations through AC elastocaloric effect \cite{Ike19.1}, which are measured through a thermocouple. Simultaneously, we measure the elastoresistance on the same bar using an amplitude demodulation technique discussed in detail elsewhere \cite{Hri18.1}. For further experimental details see Ref.\,\onlinecite{Ike19.1}.\\

\noindent{\bf Specific heat} \hspace{0.1cm}\\
The specific heat of all materials studied here has been measured using a standard relaxation time technique. For the measurements a commercially available Physical Property Measurement System from \textit{Quantum Design} has been used. Presumably, the samples experience small symmetric offset strain due to the thermal expansion mismatch to the measurement platform made out of sapphire. Neglecting the strain dependence of the transition temperatures, the specific heat is relatively weakly strain dependent. Our elastocaloric effect data is thus normalized using as-measured $C_p/T$ values. Since the conclusions drawn here are supported through elastocaloric effect data above the phase transitions, this treatment causes only negligible errors.
\newpage


\begin{acknowledgments}
The authors thank P. Walmsley, P. Massat, Y. Gallais, K. Alp Yay, J.A.W. Straquadine, and J. J. Sanchez for insightful discussions. This work was supported by the Department of Energy, Office of Basic Energy Sciences, under Contract No. DEAC02-76SF00515. M.S.I. and E.W.R. were supported in part by the Gordon and Betty Moore Foundations EPiQS Initiative through Grant No. GBMF9068. J.C.P. was supported in part by a Gabilan Stanford Graduate Fellowship and a Stanford Lieberman Fellowship as well as a Reines Distinguished Postdoc Fellowship at the National High Magnetic Field Laboratory, Los Alamos National Laboratory during the writing of the manuscript.
\end{acknowledgments}



\appendix

\section{Elastocaloric effect and elastoresistivity as a response to uniaxial stress along the tetragonal [110] axis in Co doped BaFe$_2$As$_2$ from Landau free energy analysis}
\label{sec:Landau}

In the simplest case the Landau free energy expansion for an electronic nematic phase with nematic order parameter $\Psi_{B_{2g}}$ can be written as:
\begin{equation}
\begin{split}
    \Delta{F} &= \frac{1}{2} \chi^{-1} \Psi_{B_{2g}}^2 + \frac{b}{4} \Psi_{B_{2g}}^4+ \frac{c_{66}^0}{2}\varepsilon_{B_{2g}}^2+\\
    &+\frac{d}{4}\varepsilon_{B_{2g}}^4 - \lambda \Psi_{B_{2g}} \varepsilon_{B_{2g}} - \frac{\gamma}{2}\Psi_{B_{2g}}^2\varepsilon_{B_{2g}}^2.
\end{split}
\end{equation}
Here, $b$, $c_{66}^0$, $d$, $\lambda$, and $\gamma$ are temperature independent constants, and $\varepsilon_{B_{2g}}$ is the conjugate strain component coupling bilinearly to the nematic order parameter. Note that to fourth order, also a biquadratic coupling term is allowed by symmetry, which in case $\gamma$ takes on large values, will notably tune the cross over temperature. A commonly applied form for the nematic susceptibility satisfying the conditions for a continuous phase transition is:
\begin{equation}
    \chi=\frac{1}{a_0\left(T-T^*\right)},
\end{equation}
where $a_0$ is a temperature independent constant and $T^*$ is the bare electronic transition temperature. It is important to note, that this Curie-Weiss behavior constitutes the simplest out of many continuous functions changing sign at $T^*$. Deviations from Curie-Weiss behavior for nematicity in real materials is entirely possible and even expected for instance in the presence of disorder\cite{Kuo16.1}. 
Neglecting all fourth order terms for temperatures above the phase transition, it is easy to see that the entropy associated with nematicity is proportional to $B_{2g}$ strain squared. For the $B_{\rm 2g}$ strain mediated elastocaloric coefficient at constant temperature we find
\begin{equation}
    \left(\frac{dS}{d\varepsilon_{B_{2g}}}\right)_T=-\left(\frac{d^2 \Delta F}{dT d\varepsilon_{B_{2g}}}\right)_T=\lambda^2 \frac{d\chi}{dT}\varepsilon_{B_{2g}}.
\end{equation}
Considering quasi adiabatic conditions ($dS=0$), appropriate\cite{Ike19.1} for the experiments discussed in the manuscript, the elastocaloric entropy change as a response to an oscillating stress translates into an elastocaloric temperature oscillation according to
\begin{equation}
    \left(\frac{dT}{d\varepsilon_{B_{2g}}}\right)_S=-\lambda^2 \frac{d\chi}{dT}\left(\frac{T}{C_V}\right)\varepsilon_{B_{2g}}.
    \label{eq4sup}
\end{equation}

While the discussion above captures all the essential physics, relevant for the present manuscript, it is an oversimplification for uniaxial stress experiments. Under uniaxial stress a controlled deformation is applied only along one axis (here the tetragonal [110] axis). To consider this distinct thermodynamic condition we express the free energy in terms of $\varepsilon_{\rm x^\prime x^\prime}$, which corresponds to the controlled strain along the tetragonal [110] direction, instead of $\varepsilon_{B_{2g}}$. Considering $\varepsilon_{B_{2g}}=1/2(1+\nu_{110})\varepsilon_{\rm x^\prime x^\prime}$, and neglecting contributions arising from symmetric strain (since these do not couple to $\Psi_{B_{2g}}$ linearly ) to second order we have
\begin{equation}
\begin{split}
     \Delta F &= \frac{1}{2} \chi^{-1} \Psi_{B_{2g}}^2 + \frac{c_{66}^0\left(1+\nu_{110}\right)^2}{8}\varepsilon_{\rm x^\prime x^\prime}^2 -\\ &-\frac{\lambda\left(1+\nu_{110}\right)}{2} \Psi_{B_{2g}} \varepsilon_{\rm x^\prime x^\prime}.
    \label{eq:freeE}
\end{split}
\end{equation}
Here, \[\nu_{110}=\frac{- 2c_{13}^2 + c_{11}c_{33} + c_{12}c_{33} - 2c_{33}c_{66}}{- 2c_{13}^2 + c_{11}c_{33} + c_{12}c_{33} + 2c_{33}c_{66}}\] is the in-plane Poisson ratio for uniaxial stress along the tetragonal [110] axis, which is due to the softening of $c_{66}$ temperature dependent. After minimizing the free energy with respect to the order parameter $\Psi_{B_{2g}}$ we find for the elastocaloric coefficient at constant temperature:
\begin{equation}
\begin{split}
    &\left(\frac{\partial{S}}{\partial{\varepsilon_{\rm x^\prime x^\prime}}}\right)_T =\\ &=\frac{1}{4}\lambda^2\left(\frac{\partial{\chi}}{\partial{T}}\left(1+\nu_{110}\right)^2+2\chi\left(1+\nu_{110}\right)\frac{\partial{\nu_{110}}}{\partial{T}}\right)\varepsilon_{\rm x^\prime x^\prime}\approx\\ &\approx\lambda^2\frac{\partial \chi}{\partial{T}}\frac{\left(1+\nu_{110}\right)^2}{4}\frac{T}{C_V}\varepsilon_{\rm x^\prime x^\prime}.
    \label{eq:ECET}
\end{split}
\end{equation}
To first approximation, the second term in equation\,\ref{eq:ECET} can be dropped since $2\chi{\partial\nu_{110}}/{\partial{T}}\ll{\partial\chi}/{\partial{T}}\left(1+\nu_{110}\right)$. The magnitude of the resulting effect is, for small strains (see Appendix\,\ref{sec:strainsweep}), linearly related to $\varepsilon_{\rm x^\prime x^\prime}$ (which is proportional to $\varepsilon_{B_{2g}}$) such that a sign change is observed as the applied offset strain changes from compressive to tensile. This together with a diverging temperature dependence is the hallmark for nematic fluctuations dominating the observed signal.

For the elastoresistive gauge factor $GF$, determined during our uniaxial stress experiments we expect:
\begin{equation}
\begin{split}
    GF &=2\frac{d\left(\frac{\Delta\rho}{\rho_0}\right)_{\rm x^\prime x^\prime}}{d\varepsilon_{\rm x^\prime x^\prime}}=m_{B_{2g}}^{B_{2g}}\frac{1+\nu_{110}}{2}+\\
    &+m_{A_{1g}}^{B_{2g},B_{2g}}\frac{\left(1+\nu_{110}\right)^2}{2}\varepsilon_{\rm x^\prime x^\prime}.
    \label{eq:GF}
\end{split}
\end{equation}

Here, $m_{B_{2g}}^{B_{2g}}=\lambda g_x \chi$ is the response of the antisymmetric part of the resistivity to $B_{2g}$ strain, and $m_{A_{1g}}^{B_{2g},B_{2g}}= a \lambda^2 \chi^2 +b\lambda \chi + c$ the response of the symmetric part of the resistivity to the square of the $B_{2g}$ strain. Details on the non-linear elastoresistive response are found in Ref.\,\onlinecite{Pal17.1}. We find for the temperature derivative of the gauge factor:

\begin{equation}
\begin{split}
    \frac{\partial{GF}}{\partial{T}}=&-\left(\frac{2 g_x}{\lambda \left(1+\nu_{110}\right) \varepsilon_{\rm x^\prime x^\prime}} +2\chi a +\frac{2b}{\lambda}\right)\left(\frac{\partial{S}}{\partial{\varepsilon_{\rm x^\prime x^\prime}}}\right)_T\\
    &+\left(-\frac{\lambda}{2} g_x \chi + c \varepsilon_{\rm x^\prime x^\prime} \left(1+\nu_{110}\right)\right)\frac{\partial{\nu_{110}}}{\partial{T}} +\\
    &+\frac{\lambda^2 a \varepsilon_{\rm x^\prime x^\prime}}{2}\chi \left(1+\nu_{110}\right)^2 \frac{\partial{\chi}}{\partial{T}}.
    \label{eq:dGFdT}
\end{split}
\end{equation}

The term $\frac{\lambda^2 a \varepsilon_{\rm x^\prime x^\prime}}{2}\chi \left(1+\nu_{110}\right)^2 \frac{\partial{\chi}}{\partial{T}}$ can be written in terms of $\frac{\partial S}{\partial \varepsilon_{\rm x^\prime x^\prime}}$ as

\begin{equation}
\begin{split}
    &\frac{\lambda^2 a \varepsilon_{\rm x^\prime x^\prime}}{2}\chi \left [ (1+\nu_{110})^2 \frac{\partial{\chi}}{\partial{T}} + 2\chi\nu_{110}(1+\nu_{110}) \right ] -\\
    &-\lambda^2 a \varepsilon_{\rm x^\prime x^\prime}\nu_{110}(1+\nu_{110})\chi^2\frac{d\nu_{110}}{dT}
\end{split}
\end{equation}
The term in $\left[ ... \right]$ is $-\frac{4}{\lambda^2 \varepsilon_{\rm x^\prime x^\prime}} \frac{\partial S}{\partial \varepsilon_{\rm x^\prime x^\prime}}$

Therefore, Eq.\,\ref{eq:dGFdT} turns into 

\begin{equation}
\begin{split}
    \frac{\partial{GF}}{\partial{T}}=&-\left(\frac{2 g_x}{\lambda \left(1+\nu_{110}\right) \varepsilon_{\rm x^\prime x^\prime}} +4\chi a +\frac{2b}{\lambda}\right)\left(\frac{\partial{S}}{\partial{\varepsilon_{\rm x^\prime x^\prime}}}\right)_T\\
    &+\left(-\frac{\lambda}{2} g_x \chi + c \varepsilon_{\rm x^\prime x^\prime} \left(1+\nu_{110}\right)\right)\frac{\partial{\nu_{110}}}{\partial{T}}-\\
    &- \lambda^2 a \varepsilon_{\rm x^\prime x^\prime}\nu_{110}(1+\nu_{110})\chi^2\frac{d\nu_{110}}{dT}.
\end{split}
\end{equation}

\begin{figure*}[ht]
	\centering
		\includegraphics[width=0.95\textwidth]{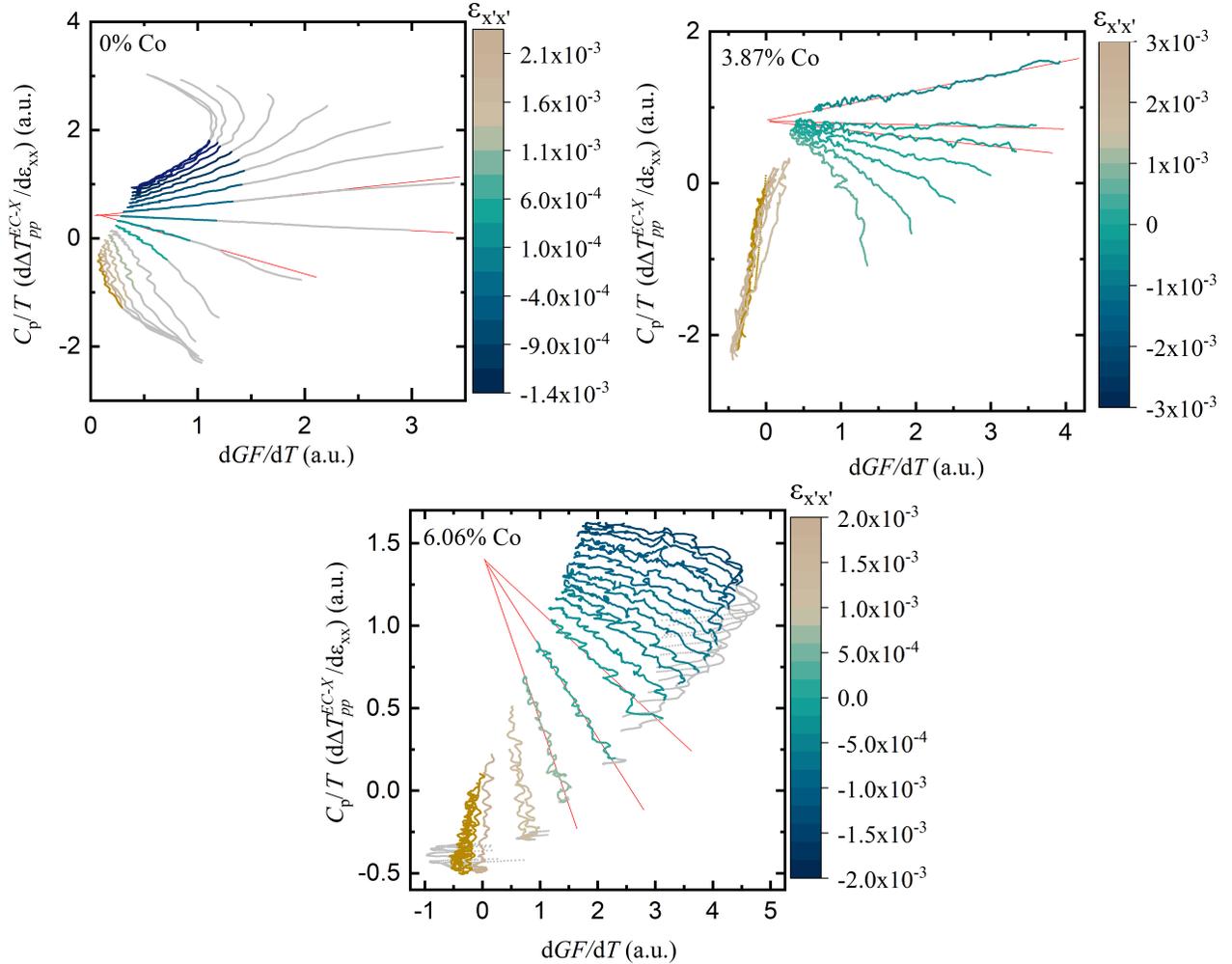}
	\caption{Under adiabatic conditions, the elastocaloric temperature oscillation multiplied by heat capacity over temperature equals the elastocaloric coefficient at constant temperature $\left(\partial{S}/\partial{\varepsilon_{\rm x^\prime x^\prime}}\right)_T$. Here, we plot this quantity for three different Co concentrations as a function of the temperature derivative of the measured gauge factor. Traces taken at different offset strains $\varepsilon_{\rm x^\prime x^\prime}$ are shown in different colors. As discussed in the text, for small strains a linear relation is expected as long as $g_x$ is not strongly temperature dependent. Data within 10\,K above the nematic cross over is shown in gray.}
	\label{FigS5}
\end{figure*}

For small strains $\varepsilon_{\rm x^\prime x^\prime}$ and small $\partial\nu_{110}/\partial{T}$, the above equation is approximated to
\begin{equation}
    \frac{\partial{GF}}{\partial{T}}=-\left(\frac{2 g_x}{\lambda \left(1+\nu_{110}\right) \varepsilon_{\rm x^\prime x^\prime}} +4\chi a +\frac{2b}{\lambda}\right)\left(\frac{\partial{S}}{\partial{\varepsilon_{\rm x^\prime x^\prime}}}\right)_T\\
       \label{eq:dGFdT2}
\end{equation}
Therefore, as long as non-linear effects to the elastoresistivity are small (at small strains and temperatures well above the nematic phase transition)
we find an approximately ($1/\left(1+\nu_{110}\right)$ is weakly temperature dependent.) linear relation between the temperature derivative of the gauge factor and the elastocaloric coefficient. This linear relation only holds as long as $g_x$ does not itself show a strong temperature dependence. This is what we find for the representative compositions described in the main text, parent ($x=0$\%), $x=$3.87\%, and $x=$6.06\% Co doped BaFe$_2$As$_2$ (see Fig.\,\ref{FigS5})
\\

\section{Elastocaloric temperature oscillations of BaFe$_2$As$_2$ as a response to $B_{2g}$ strain as a function of applied DC offset strain at fixed temperature}
\label{sec:strainsweep}
\begin{figure}[ht]
	\centering
		\includegraphics[width=0.45\textwidth]{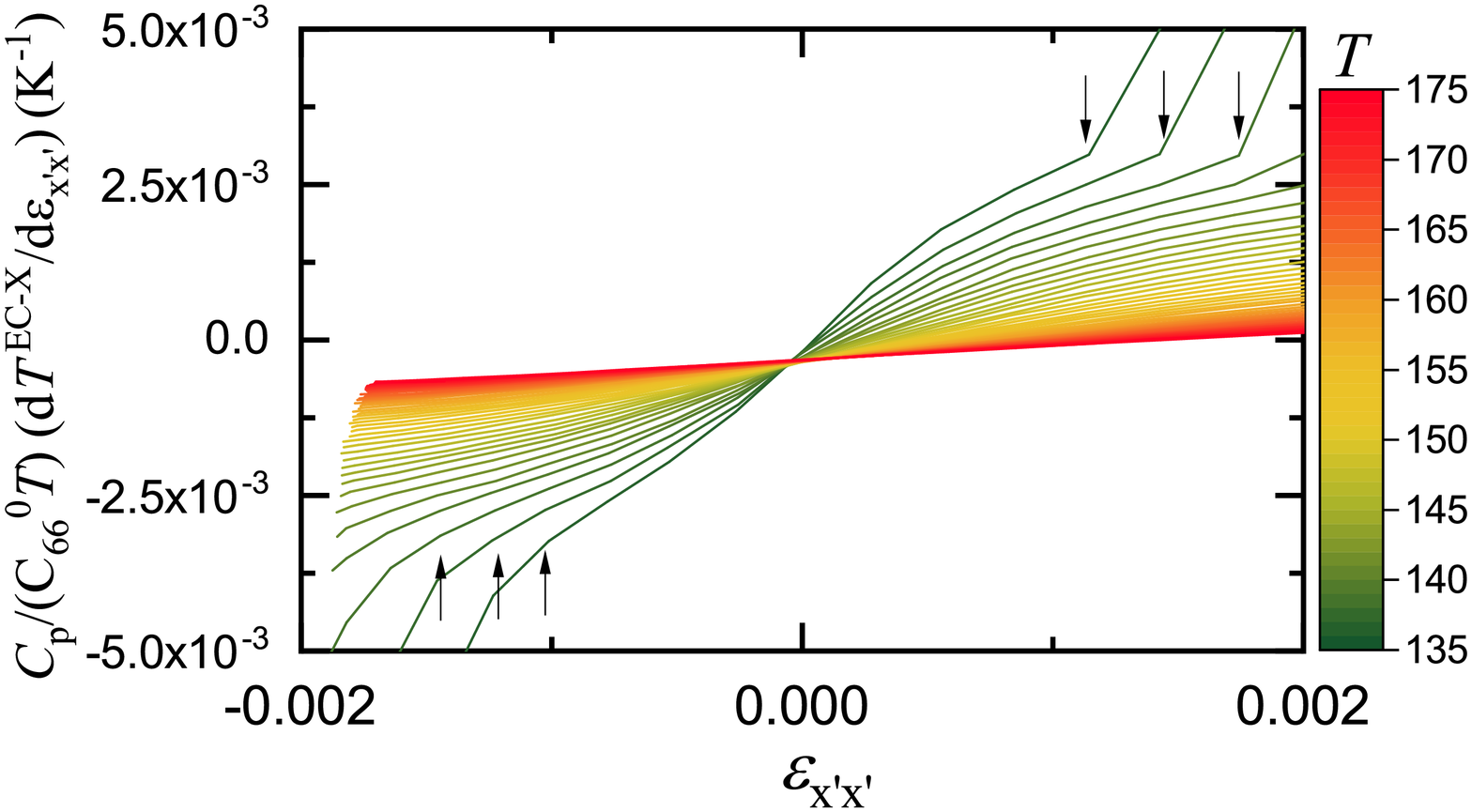}
	\caption{Elastocaloric temperature oscillation as a response to strain oscillations as a function of offset strain $\varepsilon_{\rm x^\prime x^\prime}$. Note that $\varepsilon_{\rm x^\prime x^\prime}=\varepsilon_{A_{1g}}+\varepsilon_{B_{2g}}$. The $A_{1g}$ strain derivative of the entropy is however, as discussed below, negligible for the samples studied here. The color represents different temperatures. For the traces closest to the nematic phase transition a contribution cubic in strain is observed even at relatively small strains. In addition, the kink marked by the black arrows, signifies a strain tuned nematic cross over.}
	\label{FigS4}
\end{figure}

As discussed above, considering only terms up to second order in the free energy, the elastocaloric coefficient $\left(\frac{dS}{d\varepsilon_{B_{2g}}}\right)_T$ (and thus the elastocaloric temperature oscillations under adiabatic conditions multiplied by the heat capacity over temperature) is linearly related to $B_{2g}$ offset strain. In the presence of a large biquadratic coupling between $\varepsilon_{B_{2g}}$ and the nematic order parameter the next higher order contribution to the elastocaloric effect is cubic in strain. This is what we observe for parent BaFe$_2$As$_2$ (see Fig.\,\ref{FigS4}). For small strains a linear region around the strain neutral point is observed. At high temperatures, this linear regime extends to larger strains. For temperatures close to the nematic transition temperature the linear strain range is relatively narrow and notable cubic contribution is observed. In addition, the kink in the data closest to the zero strain  phase transition temperature (which occurs at 135\,K for a free standing sample) represents the strain tuned nematic crossover. In order to analyse the temperature dependence of the $B_{2g}$ strain induced elastocaloric effect, the linear response is extracted from our data using linear fits within a small strain window around the strain neutral point.

\section{Heat capacity of BaFe$_2$As$_2$}
\label{sec:HC}
\begin{figure}[ht]
	\centering
		\includegraphics[width=0.48\textwidth]{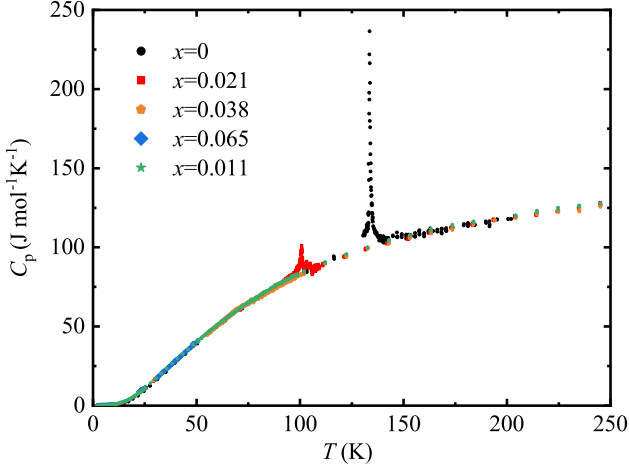}
	\caption{Heat capacity of Ba(Fe$_{1-x}$Co$_x$)$_2$As$_2$ with different Co concentration $x$. In order to eliminate the large uncertainty in the measured mass, the data has been scaled to match room temperature data of published\cite{Chu09.1} results.}
	\label{Fig:HC}
\end{figure}

The elastocaloric coefficient at constant temperature $\left(\partial{S}/\varepsilon_{\rm x^\prime x^\prime}\right)_T$ can be calculated from the measured elastocaloric temperature oscillation through a product with the heat capacity over temperature. We thus have measured the heat capacity for a selection of compositions of the Co doped BaFe$_2$As$_2$ series (see Fig.\,\ref{Fig:HC}). Fur the purposes of this manuscript we are mostly interested in the temperature dependence of $C_p/T$ above the nematic phase transition. As is obvious from the data, the heat capacity is, except for the critical contributions, weakly composition dependent. Except for parent BaFe$_2$As$_2$, for which we have used the measured data as is, we have created a generic $C_p/T$ curve from the data shown in Fig.\,\ref{Fig:HC} ignoring the critical contributions, and applied it to the measured elastocaloric effect data above the nematic phase transition temperature.

\section{Out of phase elastocaloric effect in parent BaFe$_2$As$_2$}
\label{sec:OutofPhase}

\begin{figure}[ht]
	\centering
		\includegraphics[width=0.48\textwidth]{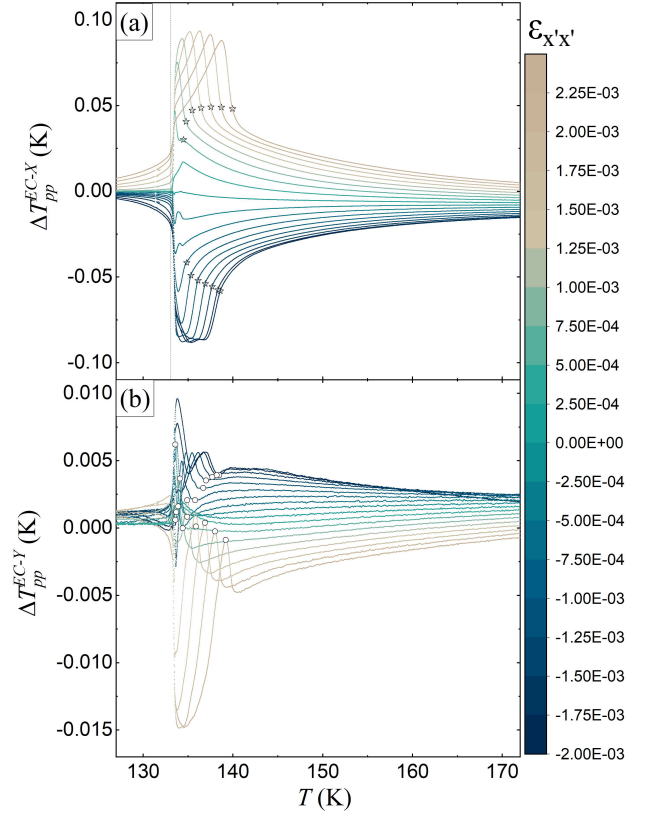}
	\caption{Peak to peak magnitude of the (a) in-phase and (b) out-of-phase elastocaloric temperature oscillation as a function of temperature. Different colors correspond to different applied offset strains. The circles in panel (b) indicate a minimum/maximum in the data, below which the out-of-phase elastocaloric effect changes rapidly across a narrow temperature window. This is likely related to dynamic effects due to domain formation.}
	\label{FigECE}
\end{figure}

Figure\,\ref{FigECE} shows the in-phase ($\Delta{T^{EC-X}}$) and out-of-phase ($\Delta{T^{EC-Y}}$) elastocaloric temperature oscillation as a function of temperature. At high temperatures, the in-phase contribution dominates, which is expected as long as the strain frequency is chosen such that the material (i) exhibits a quasi adiabatic condition and (ii) the thermometer is able to thermalize with the sample within the timescale set by the strain frequency. Details on the thermal transfer function of the materials studied here as well as on the technique can be found in Ref.\,\onlinecite{Ike19.1}. As can be seen in Fig.\,\ref{FigECE}(b), below a characteristic temperature marked by the circle symbols, the out-of-phase elastocaloric effect changes rapidly across a narrow temperature window. This is unlikely explained by a change in the thermal transfer function but is in line with dynamic effects expected due to domain formation. The characteristic temperature of this feature agrees with the feature in the in-phase data, below which the data can not be described by the expected $\left(T-T^*\right)^{-2}$ behavior (star symbols in Fig.\,\ref{FigECE}(a)) as should be expected if the order parameter has a steep temperature dependence below the transition.

\section{Elastocaloric effect of Co doped BaFe$_2$As$_2$ as a response to orthogonal $A_{\rm 1g}$ and $B_{\rm 1g}$ strain}
\label{sec:B1g}
\begin{figure}[ht]
	\centering
		\includegraphics[width=0.48\textwidth]{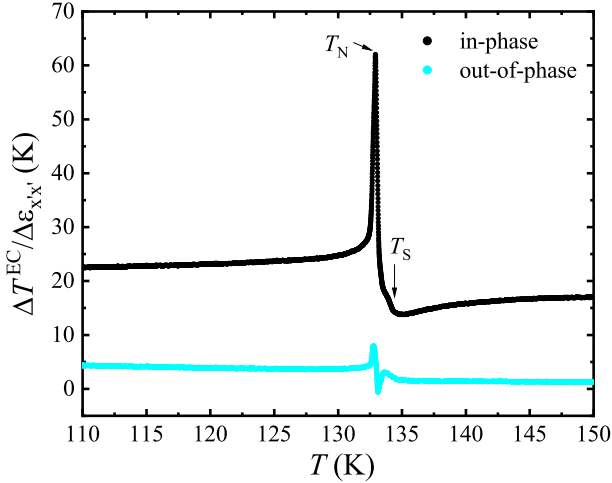}
	\caption{Elastocaloric effect of BaFe$_2$As$_2$ as a response to $B_{1g}$ / $A_{1g}$ strain determined by applying uniaxial stress along the tetragonal [100] axis. The data was taken at nominally zero offset strain.}
	\label{FigS1}
\end{figure}

\noindent Elastocaloric effect measurements on BaFe$_2$As$_2$ as a response to $A_{\rm 1g}$ and $B_{\rm 1g}$ strain (see Fig.\,\ref{FigS1}) reveal, as expected\cite{Ike19.1}, features, closely related to the critical contributions of the heat capacity. We find that, (i) the nematic and antiferromagnetic transition are clearly split by about 1K and (ii) the antiferromagnetic transition is of second order or first order involving a very small latent heat. The latent heat has to be significantly smaller than the entropy oscillations induced by the AC elastocaloric effect in order to translate into elastocaloric temperature oscillations.

\begin{figure}[ht]
	\centering
		\includegraphics[width=0.48\textwidth]{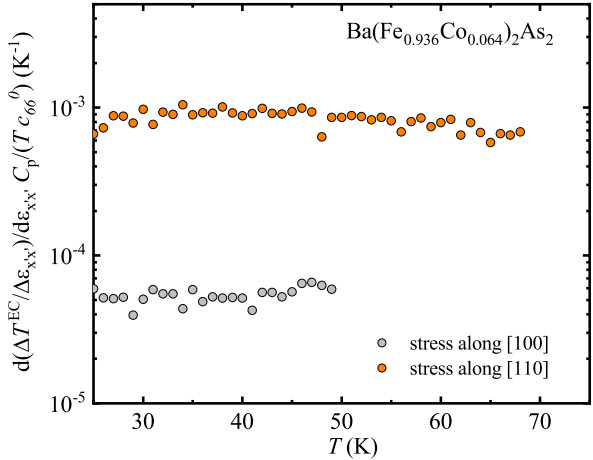}
	\caption{Strain dependence of the elastocaloric effect of 6.4\% Co doped BaFe$_2$As$_2$ multiplied by the heat capacity over temperature and scaled to units of K by division through the elastic modulus $c_{66}^0$. Note that $c_{66}^0$ and 0.5\,($c_{11}-c_{12}$) are comparable in magnitude\cite{Fuj18.1} and relatively insensitive to composition.}
	\label{FigS6}
\end{figure}

Figure\,\ref{FigS6} shows the linear strain coefficient of the elastocaloric temperature oscillation of 6.4\% Co doped BaFe$_2$As$_2$ multiplied by the heat capacity over temperature and rescaled to units of $\rm K^{-1}$ by division through the elastic modulus $c_{66}^0$ (Ref.\,\onlinecite{Fuj18.1}). We have chosen this composition for the comparison, since around optimal doping the ECE due to $B_{2g}$ strain is smallest. Under adiabatic conditions this quantity is proportional to $\left(\partial{S}/\partial\varepsilon_{\rm x^\prime x^\prime}\right)_T$. Comparing measurements applying uniaxial stress along the tetragonal [100] as well as the [110] axis, it is obvious that for the experiments focusing on stress along [110] presented in the manuscript the dominating contribution indeed stems from $B_{2g}$ strain.

\section{Determining the strain neutral point from elastocaloric effect data}
\label{sec:zeroStrain}
\begin{figure}[ht]
	\centering
		\includegraphics[width=0.48\textwidth]{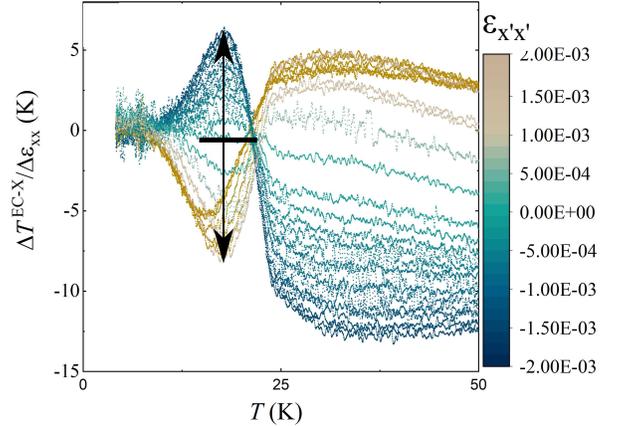}
	\caption{Elastocaloric temperature oscillation of 6.06\% Co doped BaFe$_2$As$_2$ as a function of temperature. The symbol color encodes different offset strain $\varepsilon_{\rm x^\prime x^\prime}$. At the strain neutral point, the feature for the superconducting transition is expected to be small since $\left(dT_c/d\varepsilon_{B_{2g}}\right)_{\varepsilon_{B_{2g}}=0}\approx0$. The black arrows mark the growing contribution }
	\label{FigS7}
\end{figure}

The CS100 uniaxial stress cell from Razorbill instruments is equipped with a capacitance sensor allowing for the measurement of the relative displacement of the two sides of the stress cell and thus the nominal strain in stress direction. The strain neutral point of the sample is, however, not determined from the relative displacement of the mounting plates alone, but also from its differential thermal expansion as compared to the cell body/mounting plate material. Therefore, it is desirable to find another way to determine the strain neutral point during such measurements. One way is to make use of physical properties of the sample itself. Here, we determine the strain neutral point using two different methods. For samples with low Co concentration, it is possible to determine the strain neutral point from studying the temperature dependence of the elastocaloric temperature oscillations above the nematic phase transition. If uniaxial stress is applied along the tetragonal [110] axis, the elastocaloric effect is dominated by its $B_{2g}$ strain response. This contribution is as discussed in our manuscript, linearly related $\varepsilon_{B_{2g}}$ and follows a $(T-T^*)^{-2}$ temperature dependence. A temperature sweep taken close to the strain neutral point should thus show only a very small elastocaloric effect. For low Co conentrations, the strain neutral point can be well determined using this approach (see Fig.\,\ref{FigECE}(a)). For higher Co concentrations instead of studying the temperature dependence of elastocaloric effect due to nematic fluctuations, we focus on the elastocaloric effect due to the superconducting phase transition. As we have shown earlier\cite{Ike19.1}, the elastocaloric effect in the vicinity of continuous phase transitions is proportional to the critical contribution divided by the total heat capacity (this is true unless a strain representing a conjugate field for the phase transition is applied). Importantly, the proportionality constant is the strain dependence of the critical temperature $dT_c/d\varepsilon$. If stress is applied along the tetragonal [110] axis, the leading contribution on the superconducting transition temperature comes from $B_{2g}$ strain. By symmetry, we expect a quadratic dependence of the superconducting phase transition temperature on 
$B_{2g}$ strain such that $dT_c/d\varepsilon$ should go through zero at the strain neutral point. This is what we observe (see for example Fig.\,\ref{FigS7}) for the superconducting transition of Co doped  BaFe$_2$As$_2$ when stress is applied along the tetragonal [110] axis.\\

\section{Estimating the temperature dependence of the in-plane Poisson ratio and quantitative analysis of the $B_{2g}$ elastocaloric effect of parent BaFe$_2$As$_2$}
\label{sec:Poisson}
When uniaxial stress is applied to the tetragonal [110] axis of Co doped BaFe$_2$As$_2$, the in-plane Poisson ration $\nu_{110}$ is expected to be temperature dependent and vary from about 0.3 at high temperatures to 1 at the nematic phase transition as $c_{66}^0$ softens to zero. A temperature dependent Poisson ratio causes a temperature dependent magnitude of $B_{2g}$ strain in our experiments since only the strain along the stress direction $\varepsilon_{\rm x^\prime x^\prime}$ is controlled.

\begin{figure}[ht]
	\centering
		\includegraphics[width=0.48\textwidth]{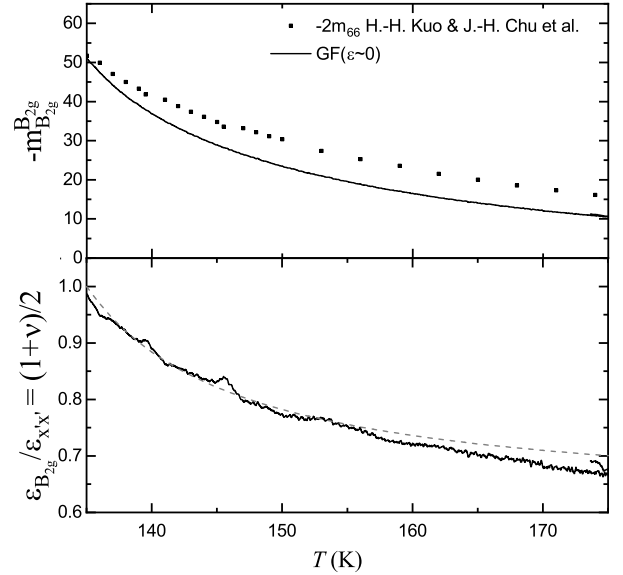}
	\caption{Comparison of $2m_{66}$ elastoresistivity and elastoresistive gauge factor determined during our uniaxial stress experiments of BaFe$_2$As$_2$. The bottom panel shows the ratio between the data shown in the top panel as solid black line. This ratio corresponds to strain ratio $\varepsilon_{B_{2g}}/\varepsilon_{\rm x^\prime x^\prime}$. The dashed gray line is the theoretically expected strain ratio, taking into account a full softening of $c_{66}$ as well as a Curie-Weiss like nematic susceptibility.}
	\label{FigS2}
\end{figure}

\begin{figure}[ht]
	\centering
		\includegraphics[width=0.48\textwidth]{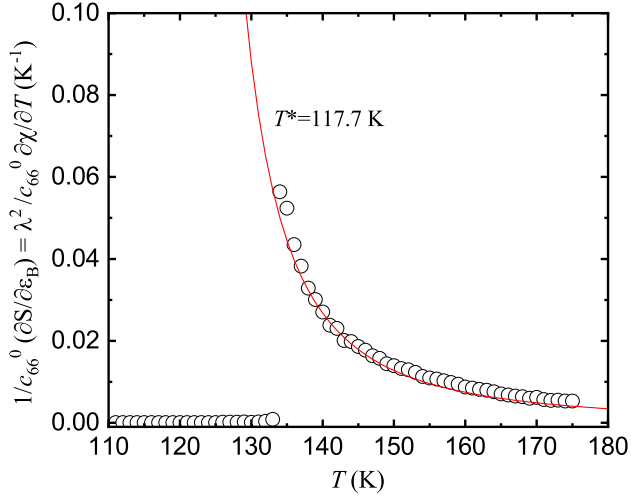}
	\caption{Strain dependence of the elastocaloric temperature oscillations multiplied by heat capacity over temperature and scaled to units of $\rm K^{-1}$ through dividing by $c_{66}^0$. Under adiabatic conditions this quantity correponds to $1/c_{66}^0 \left(\partial{S}/\partial\varepsilon_{B_{2g}}\right)_T$. The red line shows a fit to $0.77 (T_S-T^*)/(T-T^*)^2$. The factor 0.77 takes into account that only a fraction of the elastocaloric temperature oscillation is detected\cite{Ike19.1}. Our fit reveals a Weiss temperature (since the nematic transition temperature is fixed to the value observed for freestanding samples, $T^*$ is the only free parameter) of 117.7\,K.}
	\label{FigS3}
\end{figure}

This is a thermodynamic condition different from controlling $B_{2g}$ strain (see Appendix\,\ref{sec:Landau}) and yields a slightly more complicated expression for the elastocaloric temperature oscillation as a response to oscillating strain $\varepsilon_{\rm x^\prime x^\prime}$. The in-plane Poisson ratio as well as the ratio between $B_{2g}$ strain and $\varepsilon_{\rm x^\prime x^\prime}$, which corresponds to $1/2\left(1+\nu_{110}\right)$, can be calculated assuming a softening of $c_{66}^0$ according to a Curie-Weiss behavior of the nematic order parameter. Here we show, that a measurement of the elastoresistive gauge factor $GF=d\left(\Delta\rho/\rho_0\right)_{\rm x^\prime x^\prime}/d\varepsilon_{\rm x^\prime x^\prime}$ allows for estimating the in plane Poisson ratio as function of temperature when compared to the $2m_{66}$ elastoresistivity component available in literature\cite{Kuo16.1}. The top panel in Fig.\,\ref{FigS2} shows a comparison of our measured gauge factor to published\cite{Kuo16.1} $2m_{66}$ elastoresistivity data. The gauge factor is, in the absence of non-linear effects in the elastoresistivity, expected to match $2m_{66}$ if the Poisson ratio takes on a value 1, such that $\varepsilon_{\rm x^\prime x^\prime}=\varepsilon_{B_{2g}}$ (since $\varepsilon_{yy}=-\varepsilon_{\rm x^\prime x^\prime}$ if $\nu_{110}=1$, and  $\varepsilon_{B_{2g}}=0.5\left(\varepsilon_{\rm x^\prime x^\prime}+\varepsilon_{yy}\right)$). Our observation is that this is approximately the case at the nematic phase transition temperature. The ratio between the elastoresistivity data shown in the top panel of Fig.\,\ref{FigS2} gives an estimation for $\left(1+\nu_{110}\right)/2$, shown in bottom panel. Our estimation shown as black line is compared to the theoretically expected strain ratio, based on full softening of $c_{66}$ as well as a Curie-Weiss like nematic susceptibility. Our estimation agrees well with the prediction and we thus use it to extract the $B_{2g}$ strain dependence of the elastocaloric effect. Figure\,\ref{FigS3} shows this quantity multiplied by the heat capacity over temperature and scaled to units of $\rm K^{-1}$ through division by $c_{66}^0$. Within quasi adiabatic conditions, this quantity is equal to $1/c_{66}^0 \left(\partial{S}/\partial\varepsilon_{B_{2g}}\right)_T$. As discussed in detail in the manuscript, we expect this quantity to be proportional to $(T-T^*)^{-2}$. The proportionality constant also depends on the Weiss temperature $T^*$ and is given by $T_S-T^*$. Since the nematic transition temperature is obvious from our data,  the Weiss temperature is very well determined through a power-law fit. Such a fit is shown for the data on BaFe$_2$As$_2$ as red line in Fig.\,\ref{FigS3}. It is important to note that for the fit, we added 0.77 as factor in order to take into account, as has been pointed out earlier\cite{Ike19.1}, that only a fraction of the total elastocaloric temperature oscillation is detected. Our fit reveals a Weiss temperature of 117.7\,K, which is in excellent agreement with a $T^*$ found through elastoresistivity\cite{Kuo16.1}. This analysis heavily relies on the correction for the temperature dependence in $B_{2g}$ strain. Since for larger Co dopings, non-linear elastoresistive effects significantly complicate the comparison of the measured uniaxial gauge factor to $2m_{66}$ data, we here restrict the quantitative analyses to parent BaFe$_2$As$_2$.

\cleardoublepage
\newpage
\noindent{\bf REFERENCES}

%

\end{document}